\documentclass[journal]{IEEEtran}
\usepackage{booktabs} % For formal tables
\usepackage{graphicx,url, amsmath, arrayjobx}
\usepackage{float}
\usepackage{amsthm, amssymb}
\usepackage{soul}
\usepackage{color}
\usepackage{algorithmic}
\usepackage{caption}
\usepackage{setspace}
\usepackage{subfigure}
\usepackage{subfigure}

\usepackage[linesnumbered, boxed, commentsnumbered, ruled]{algorithm2e}
\usepackage{amsfonts}

\usepackage[ruled]{algorithm2e} % For algorithms

\SetAlFnt{\small}
\SetAlCapFnt{\small}
\SetAlCapNameFnt{\small}
\SetAlCapHSkip{0pt}
\IncMargin{-\parindent}
\SetKwProg{Fn}{Function}{}{}
\title{A Security and Performance Driven Architecture for Cloud Data Centers}
\author{Muhamad Felemban, Anas Daghistani, Yahya Javed, Jason Kobes, and Arif Ghafoor}

\begin{document}
\maketitle

\section*{Abstract}
With the growing cyber-security threats, ensuring the security of data in Cloud data centers is a challenging task. A prominent type of attack on Cloud data centers is data tampering attack that can jeopardize the confidentiality and the integrity of data. In this article, we present a security and performance driven architecture for these centers that incorporates an intrusion management system for multi-tenant distributed transactional databases. The proposed architecture uses a novel data partitioning and placement scheme based on damage containment and communication cost of distributed transactions. In addition, we present a benchmarking framework for evaluating the performance of the proposed architecture. The results illustrate a trade-off between security and performance goals for Cloud data centers. 

\section{Introduction}
\begin{figure}[t!]
  \centering
  {  \includegraphics[scale=0.3]{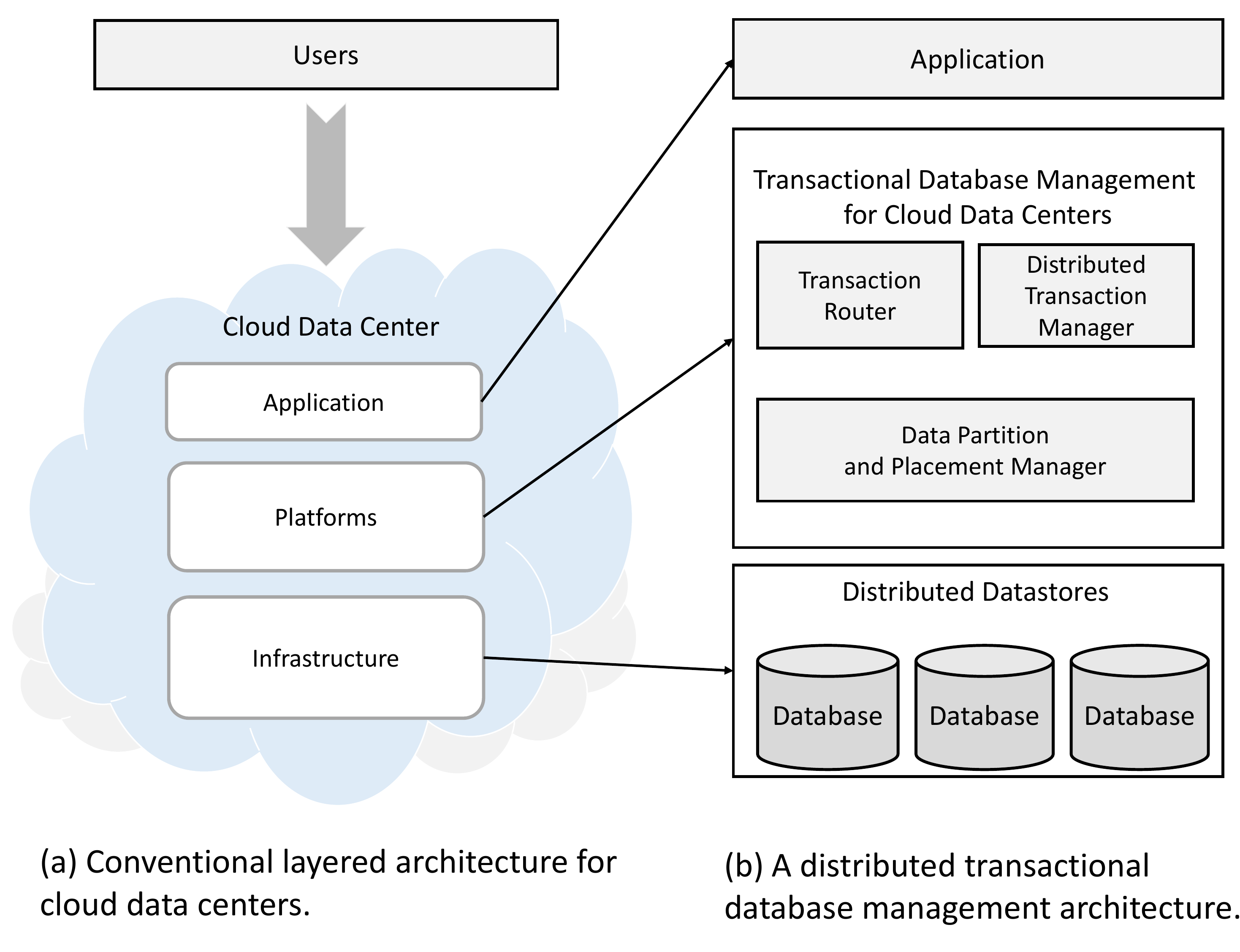}}
  \caption{CDC and distributed transactional database architectures.}
  \label{fig:allcdc}
\end{figure}

The rapid growth of data volume driven by IoT, social networks, and other data-intensive applications poses numerous performance and security challenges in terms of real-time data stores and processing \cite{ali2016security}. Cloud computing has emerged as a leading infrastructure that provides a pay-per-use access to a shared pool of resources. Numerous research efforts have aimed at designing scalable Cloud Data Centers (CDCs) for hosting multi-tenant applications \cite{mann2015allocation}. Figure \ref{fig:allcdc}(a) depicts the conventional three layered architecture of a CDC. These layers include: Application, Platform, and Infrastructure layer. The application layer provides access to various services to the end-user. The platform layer consists of operating systems and software frameworks including database systems for supporting Cloud applications. The infrastructure layer consists of shared physical resources, e.g., CPUs, memory, and data storage. 

One of the main challenges in designing transactional databases for CDCs is the partitioning and placement of tenants data across a cluster of data stores. Efficient solution to this problem can result in an increase in the availability and reliability of the CDC \cite{kumar2014sword,mann2015allocation}. This is especially true for multi-tenant applications that may serve large volume of transactions. An example of a transactional database designed for Cloud environments is \textit{ElasTras} \cite{das2013elastras}. Figure \ref{fig:allcdc}(b) depicts the architecture of a typical distributed transactional database, which consists of three components: a distributed transactions manger (DTM), a transactions router, and a data partition manger. The DTM is responsible for managing the commit protocol of the distributed sub-transactions. The transaction router maps each sub-transaction to an appropriate database that contains the required data. The partition manger is responsible of managing the distributed storage solutions, e.g., Hadoop Distributed File System (HDFS), by creating and placing the partitions across the distributed data store. Several solutions have been proposed to address the problem of data partitioning and placement for distributed Cloud data stores. These solutions are primarily focused on achieving various performance criteria such as latency and transaction throughput. \cite{kumar2014sword,turcu2016automated}. 

On the other hand, the multi-tenancy feature provided by a Cloud infrastructure poses unique security risks due to data leakage, virtual machine escape,  and side channel attacks. Such risks arise from sharing of physical resources among tenants. Another prominent intrusion attack in transactional databases is the \textit{data tampering attack} that aims at modifying confidential data with incorrect values \cite{puthal2016threats}. Due to multi-tenancy feature of CDCs, a data tampering attack can cause catastrophic cascading failures and performance degradation as a result of application interoperability, data dependency, and data sharing among tenants. Such attacks can be manifested by exploiting vulnerabilities in the application, e.g., SQL injection, privilege escalation, and virtual machine escape. Table \ref{table:synpar} lists some known Cloud vulnerabilities along with their scores as stipulated by the Common Vulnerability Scoring System (CVSS)\footnote{https://www.first.org/cvss/}. Exploitation of these vulnerabilities constitutes data tampering attacks in CDCs. Several solutions have been proposed to address the risk of data leakage including data isolation and enforcement of some access control mechanisms \cite{almutairi2012distributed}. However, no solution has been proposed to mitigate the risk of intrusion attacks on transactional databases for CDC. 

\begin{table*}[ht!]
\centering
\caption{Example of vulnerabilities in CDCs.}
\begin{tabular}{|p{3cm}|p{1cm}|p{13cm}|} \hline  
Vulnerability & CVSS Score & Description \\ \hline \hline 
CVE-2017-7546 & 9.8 & PostgreSQL vulnerability that allows adversary to gain access to database with empty password   \\ \hline
CVE-2018-1058 & 8.8 & PostgreSQL vulnerability that allows adversary to execute codes in with the permissions of superuser \\ \hline
CVE-2015-7502 & 5.1 & Operating systems vulnerability that allows adversary to obtain sensitive data and gain privilege \\ \hline
CVE-2018-5985 & 9.8 & SQL injection vulnerability in LiveCRM SaaS cloud component \\ \hline
CVE-2016-9994 & 7.1 & SQL injection vulnerability in IBM Cloud \\ \hline
\end{tabular}
\label{table:synpar}
                    \vspace{-3 mm}
\end{table*}

In this article, we propose a security-driven architecture for transactional database for CDCs. Development of this architecture entails reengineering the CDC architecture by including a real-time Adaptive Intrusion Management System (AIMS) that provides intrusion detection, response and recovery mechanisms for the database. AIMS uses an adaptive access and admission control mechanism that responds to intrusion attacks by selectively blocking segments of data that are affected by these attacks. Intrusion attacks are manifested in the form of malicious transactions. The proposed architecture uses a security-driven and performance-oriented data partitioning and placement strategy across Cloud data stores. We model the data partitioning and placement requirement as an optimization problem with combined performance and security objectives. In addition, we present a novel malicious transaction benchmark for AIMS to illustrate its resiliency against various attack scenarios. We present the evaluation results to highlight the viability  of the proposed architecture and illustrate a trade-off between security and performance in the context of reengineering CDCs.

\section{Related Work}

Several solutions have been proposed to prevent and mitigate the effect of intrusion attacks on transactional databases. One solution is to employ an Intrusion Detection System (IDS) with the objective of an IDS is to monitor and detect illegal accesses and malicious actions in transactional databases \cite{tan2014enhancing,sallam2017dbsafe}. However, an IDS can miss the detection of an attacks and is not designed to repair the damage caused by late detection of attacks. Therefore, an IDS is often integrated with response and recovery mechanisms to alleviate the damage \cite{ammann2002recovery}. In \cite{chandra2011intrusion}, a mechanism to recover from intrusion attacks for web-applications by rolling back the database and replaying subsequent legitimate actions to correct the state of the database is proposed. In \cite{pardal2017rectify}, the authors propose an intrusion recovery tool for database-driven applications running on Platform-as-a-Service Clouds. AIMS can be used as a middle-layer between the transactional database and the application to perform automatic intrusion response and recovery independently from the running applications. The most relevant work close to AIMS is given in \cite{bai2009data,ammann2002recovery} that proposes an online damage tracking and quarantine mechanism to increase the survivability of single-tenant databases. The authors propose a multi-pass recovery procedure to ensure that all the corrupted data objects are recovered. However, in scenarios with high dependency among transactions in a workload, the recovery procedure can take a long time and hence can degrade the overall performance. AIMS employs an admission control mechanism that momentarily suspends running transactions and performs the recovery procedure in a single-pass. 

Previous work on data partitioning and placement has been focused on improving the availability, reliability, and scalability, but not on the security aspect of shared-nothing distributed transactional databases. For example, in \cite{golab2014distributed} authors study the problem of data placement strategies that minimize the data communication cost incurred by distributing data across clusters of servers. In \cite{turcu2016automated}, authors propose an automatic data partitioning methodology for distributed transactional memory systems. In \cite{kumar2014sword}, authors propose a partitioning and placement technique across large number of machines in order to minimize the number of distributed transactions. However, in this article the proposed architecture of CDC addresses the challenge of data partitioning and placement with respect to both performance and security considerations.

\section{Security-Driven Reengineering Design of Cloud Data Center}

Figure 2(a) depicts a generic architecture of AIMS, which provides intrusion detection, response and recovery mechanisms for the database. In the following we briefly elaborate these mechanisms. Subsequently, in order to reengineer the design of the CDC architecture in Figure 1(b), we integrate AIMS functionalities in this architecture. The integration entails the identification of new components and possible re-designing  some of the existing components of Figure 1(b). In Figure 2(b) the integrated architecture of CDC is depicted. This architecture provides a security-driven and performance-oriented data partitioning and placement, and distributed intrusion management across cluster of servers for CDC. Discussion on AIMS and each component of the new architecture is given in the following sections.
%
%The proposed security-driven architecture for transactional database in cloud data centers includes two components. The first component is AIMS, which provides intrusion detection, response and recovery mechanisms for the database. The second component is an integrated transactional database architecture with AIMS. The modified architecture addresses provides a security-driven and performance-oriented data partitioning and placement, and distributed intrusion management across cluster of servers on the cloud. Elaborate discussion on each component is provided in the following sections. 

\subsection{Adaptive Intrusion Management System}

The objective of AIMS is to mitigate the damage caused by intrusion attacks on transactional databases. We consider transaction-level intrusion attacks on the database. A transaction, $t_m$, is malicious if it tampers the database by updating data objects with incorrect values. In this context, a malicious transaction corrupts the data by launching an attack. A transaction, $t_a$, is affected if it directly (or indirectly) depends on a malicious or an another affected transaction. Two transactions, $t_i$ and $t_j$, are dependent if $t_j$ reads an object that has been updated by $t_i$. Malicious and affected transactions are invalid transactions, which, if executed, take the database into an invalid state. Consequently, the integrity and availability of the CDCs can be affected. 

\begin{figure}[t!]
  \centering
            \subfigure[Generic AIMS Architecture.]{\label{fig:ims_arch}  \includegraphics[scale=0.3]{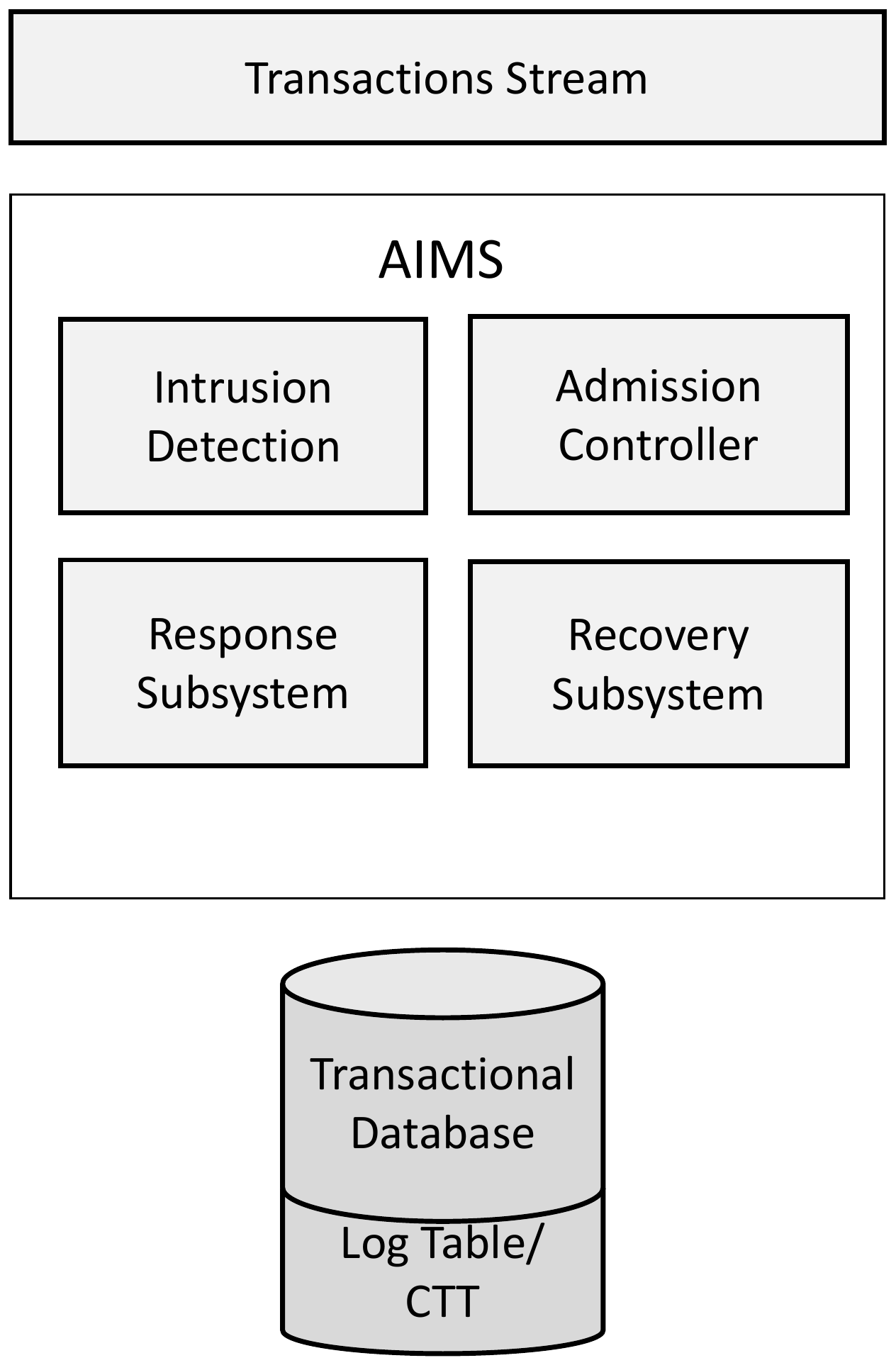}}
  \subfigure[Security-driven distributed transactional database architecture.]{\label{fig:daims}  \includegraphics[scale=0.3]{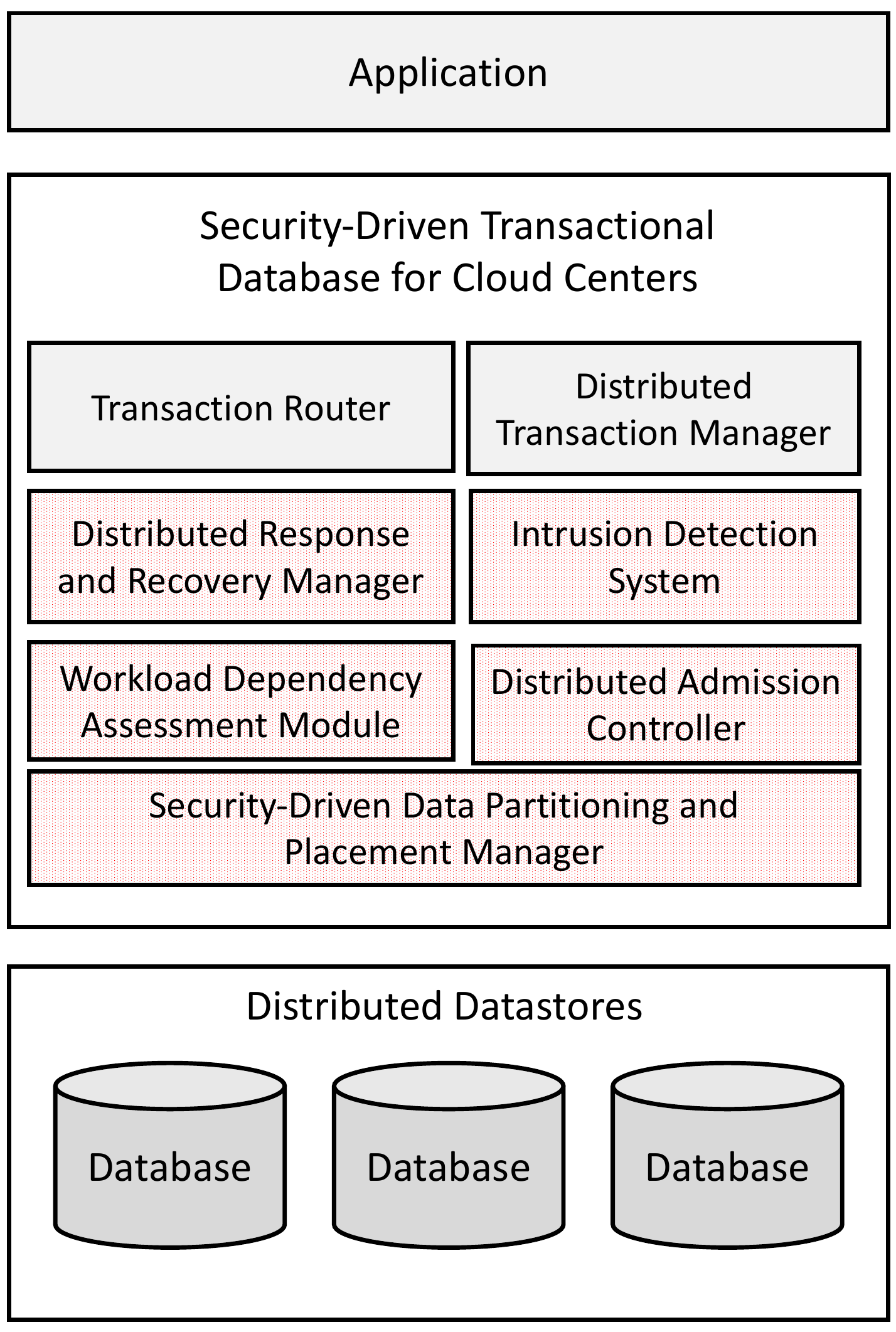}} 
%     \subfigure[Number of affected transactions.]{\label{fig:aff3}\includegraphics[scale=0.3]{figures/gnuplot/jointing/aff.eps}}%
%    \subfigure[Communication cost.]{\label{fig:cost3}\includegraphics[scale=0.3]{figures/gnuplot/jointing/cost.eps}}
  \hspace{0.2cm}
  \caption{Reengineering CDC by integrating AIMS with the distributed transactional database architecture.}
  \label{fig:daims_aims}
\end{figure}

A generic architecture of AIMS, as depicted in Figure \ref{fig:ims_arch}, contains four components: the IDS, the Admission Controller, the Response Subsystem, and the Recovery Subsystem. We assume that existing IDSs can be integrated with AIMS. Further discussion about IDS is not be provided in this article; we refer interested readers to \cite{tan2014enhancing,sallam2017dbsafe}. In the following subsections, we discuss each component in detail. The pseudocode of the procedures performed by each component is listed in Algorithm \ref{alg:tms}.

\subsubsection{Admission Controller}
\label{sec:ac}

The main functionality of the admission controller is to regulate the execution of the transactions. For this purpose, the admission controller checks if the transaction is requesting to access corrupted objects, which are stored in the Corrupted Objects Table (COT). In particular, the admission controller extracts the read/write set, $RW$, of the transaction. If $RW \cap COT$ is not empty, then the transaction is blocked until the requested objects are recovered and released from COT by the recovery subsystem as discussed below.

\subsubsection{Response Subsystem}

The objective of the response subsystem is to provide an initial evaluation of the damage caused by a malicious transaction $t_m$ in order to prevent incoming transactions from further spreading the damage. In particular, IDS alerts the response subsystem when $t_m$ is detected as malicious. Subsequently, the response subsystem collects the time information about $t_m$, i.e., its commit timestamp $t_m^c$ and its detection timestamp $t_m^d$. The response subsystem marks all the objects that have been updated during the period $t_m^c$ and $t_m^d$ by adding them to COT. 

\subsubsection{Recovery Subsystem}
\label{sec:REC}

The objectives of recovery subsystem are: 1) to identify the \textit{correct} and \textit{complete} set of affected transactions, and 2) to execute the compensating transactions accordingly. The set of affected transactions is correct and complete if and only if the set contains no transactions that are falsely identified as affected and contains every affected transactions caused by the attack. In order to identify this set, the recovery subsystem temporarily blocks new transactions to prevent them from reading any undiscovered corrupted objects. This subsystem performs blocking operations based on a locking algorithm. When the lock is acquired by the recovery subsystem, the admission controller blocks all the incoming transactions until the recovery subsystem releases the lock. During the time in which the lock is acquired by the recovery subsystem, the correct and complete set of affected transactions is identified. Subsequently, uncorrupted objects are released from COT. 

The recovery procedure is performed in two phases. First, the recovery subsystem executes compensating transactions to undo the effect of the malicious and affected transactions. The recovery subsystem uses the transactions log table to find the correct version of the corrupted objects. In particular, the corrupted objects are updated with the values of the most recent versions before the execution of the malicious transaction. Second, the recovery subsystem executes compensating transactions to re-execute each transaction in the set of affected transactions. The information required to re-execute the transactions is maintained in the transactions log. At the end of the recovery procedure, the recovered data objects are removed from COT. Also, the admission controller is signaled to resume the admission of the blocked transactions.

\begin{algorithm}[t!]
\small
\SetNlSty{normal}{}{.}
\Fn{Admission Controller ($t_i$)}{
\tcc{Invoked when a transaction $t_i$ is executed}
$RW_{t_i} \leftarrow$ the read/write objects set $t_i$\\
\While{$RW_{t_i} \cap COT$}{
Block $t_i$
}
Execute $t_i$\\
}
\Fn{Response Subsystem ($t_m, t_m^d$)}{
\tcc{Invoked when IDS detects a transaction $t_m$ as malicious}
$t_m^c$ $\leftarrow$ get commit time of $t_m$\\
\For{objects updated between $t_m^c$ and $t_m^d$}{
Add object to $COT$
}
}
\Fn{Recovery Subsystem ($t_m$)}{
\tcc{Invoked when the response subsystem is terminated}
\textbf{Block new transactions}\\
$AT \leftarrow$ Find the complete and correct set of affected transaction\\ 
\textbf{Resume new transactions}\\
Remove uncorrupted objects from COT\\
Undo $t_m$ and all $t \in AT$ \\
Redo all $t \in AT$ \\
}
\caption{AIMS procedures}	
\label{alg:tms}
\end{algorithm}

\subsection{Security-Driven Transactional Database for Cloud Data Centers}

Figure \ref{fig:daims} shows the modified security-driven architecture of transactional database for CDCs. The key modification of the proposed architecture is the redesiging of the Data Partitioning and Placement Manager (DPPM) in Figure \ref{fig:allcdc}(b) to provide damage containment in the presence of an attack. The new security-driven DPPM uses information about inter-transaction dependency to generate an optimal data partitioning and placement plan that minimizes the communication cost and curtails the propagation of damage among tenants. This information is provided by the Workload Dependency Assessment (WDA). The objective of WDA is to analyze and capture the inter-transaction dependencies in the workload. Note, we only consider data partitioning and placement plans with no data replication. As a result, a distributed transaction is divided into sub-transactions, where each sub-transaction is executed in different data stores. In addition, four new components are added to the architecture in Figure \ref{fig:allcdc}(b) to perform the functionalities of AIMS, namely, the Distributed Response and Recovery Manager (DRRM), the Distributed Admission Controller (DAC), the WDA, and the IDS. The objective of the DRRM is to coordinate the response and recovery procedures across the distributed data stores. The objective of the DAC is to coordinate the admission of incoming transactions based on the status of aggregated COT of each data store. Note, the IDS, the DAC, and the WDA operate at the transaction level, i.e., before a transaction is divided to multiple sub-transactions, which are routed to their respective data store. On the other hand, the DRRM operates at the sub-transaction-level. In essence, the DRRM manages the response and recovery procedures to recover the corrupted objects damaged by executing the sub-transactions in  each data store. In the next section, we discuss the security-driven data partitioning and placement scheme deployed by DPPM. 

\section{Reengineering Challenge: Data Partitioning and Placement for Cloud Data Center}

\begin{figure}[t!]
  \centering \includegraphics[scale=0.35]{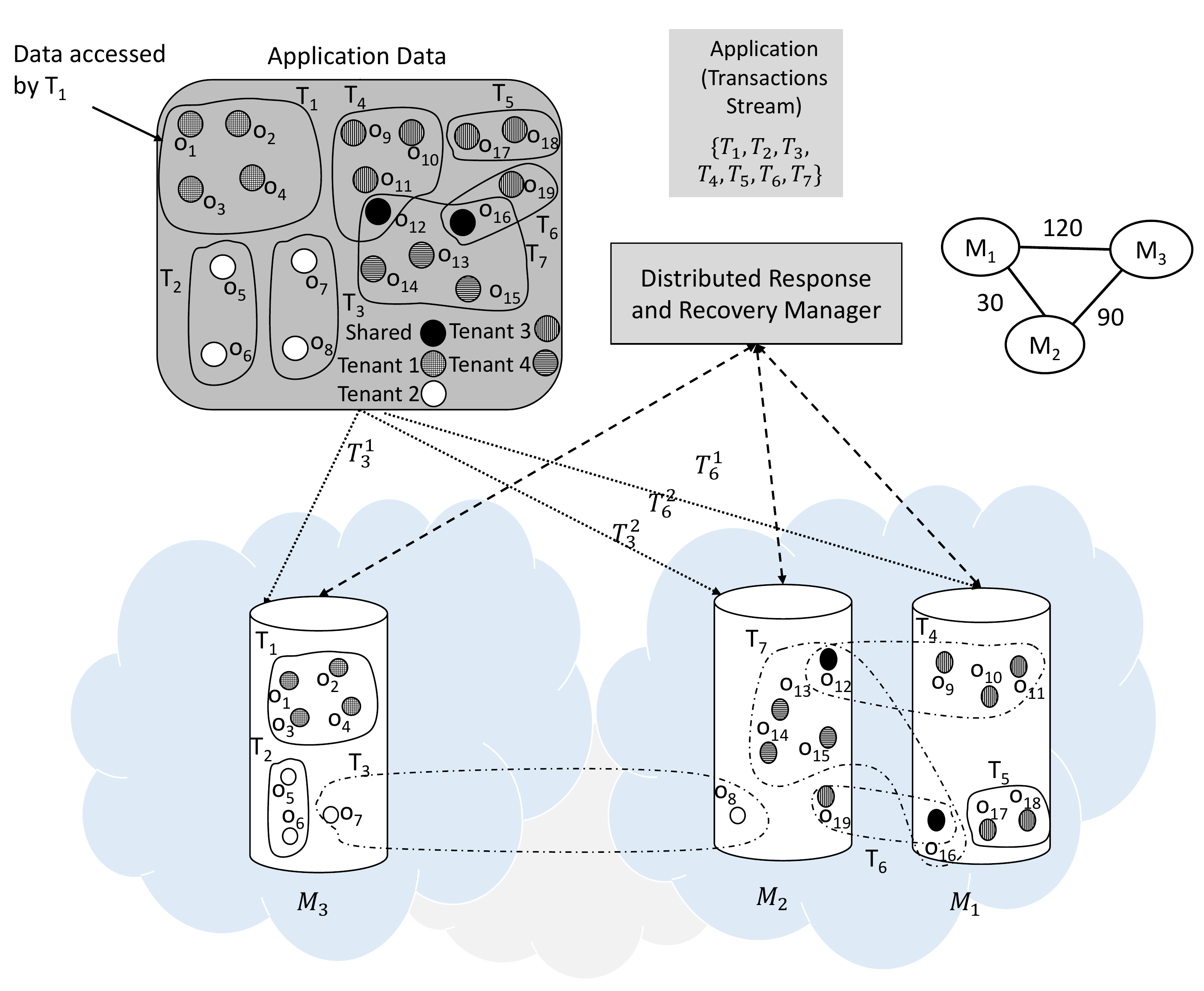}
  \hspace{0.2cm}
  \caption{Pictorial example of data partitioning and placement in CDCs. Each cylinder is a data store representing a data partition.}
  \label{fig:example}
\end{figure}

We perceive that the distributed data stores of CDC can provide a better damage containment strategy using an intelligent data partitioning and placement scheme. Consider for example a CDC with three data stores, i.e., $M_1$, $M_2$, and $M_3$, in Figure \ref{fig:example}. Let $D$ be the set of objects in the multi-tenant workload, where $D_i$ denotes the data objects of the $i^{th}$ tenant. Note, $D = \bigcup_{i} D_i \cup D_S$, where $D_S$ is the set of shared data objects among the tenants. Let the set of transactions executed by the tenants on $D$ be $T$, which are executed in the following order $\{ T_1,T_2,T_3,T_4,T_5,T_6,T_7\}$. A transaction can be classified into two categories: \textit{single-tenant transactions} and \textit{multi-tenant transactions}\footnote{https://docs.microsoft.com/en-us/azure/sql-database/}. Single-tenant transactions access data objects belonging to a single tenant, whereas multi-tenant transactions access shared objects among tenants and are denoted with $T'$. For example, $T'=\{ T_4,T_6,T_7\}$. Note, $T' \subseteq T$. Refer to the diagram in the top-left corner of Figure \ref{fig:example}. If $T_6$, executed by Tenant 3, is malicious, then the data of Tenant 4 will be corrupted if $T_7$ is executed without waiting for the decision of the IDS. A plausible solution to contain the damage is by employing a partitioning and placement scheme that splits and places the data accessed by $T_6$ to $M_1$ and $M_2$. Consequently, $T_6$ becomes a distributed transaction as shown in diagram in the bottom-right corner of Figure \ref{fig:example}. Since the sub-transactions of a distributed transactions need to be coordinated at the time of commit, the transaction is held until the IDS declares if it is a malicious or a benign transaction. Note, a naive solution for damage containment is to wait for the IDS decision after the execution of each transaction. However, this can substantially reduce the transaction throughput for the CDC. 

The goal of the data partitioning and placement scheme is to help in containing the damage caused by malicious transactions. However, data partitioning can incur some performance degradation as a result in increase in the cost of communication among distributed data stores. This communication is incurred due to the delay among data stores \cite{kumar2014sword}. Therefore, we propose a scheme that jointly optimizes the security and performance goals in terms of data partitioning and placement across data stores. Subsequently, we can formulate the partitioning and placement scheme as a dual-objective optimization problem with the joint objective to minimize the communication cost while ensuring damage containment. Alternatively, the problem can be transformed into a single-objective optimization problem to minimize the communication cost of distributed transactions. In this case, the damage containment is modeled as a constraint function in the optimization problem that forces a transaction with shared data to span multiple data stores. Let $S_t$ be the span of transaction $t$, i.e., the set of data stores that $t$ is accessing through its sub-transactions. The communication cost of $t$ is denoted as $C(S_t)$ and is given as the following. 

\begin{equation}
C(S_t) = \max_{\underset{u,v \in S_t, u \neq v}{e \in (u,v)}} \ell(e)
\end{equation}
where $\ell(e)$ is the value of the label on edge $e$. In other words, the communication cost of transaction $t$ accessing the set of partitions (i.e. data stores) represented by $S_t$ is the maximum delay incurred by the communication between any pairs of data stores. For example, the communication cost of $T_3$ is 90, since $S_{T_3}=\{M_1,M_2\}$ as shown in the graph in the top-right corner in Figure \ref{fig:example}. Accordingly, the optimization problem for the partitioning and placement scheme of the reengineered CDC can be given as follows. 

\begin{align}
& \underset{P \in \mathcal{P}}{\text{Minimize }} 
& &  C(S_t) \\ \nonumber
& \text{subject to} \\ 
& & & |S_{t'}| > 1 \ \ \ \forall t' \in T'  \label{eq:dist}\\ 
& & & |P_i| \leq Capacity(M_i) \ \ \ \forall i =1, \dots, n \label{eq:capcon} 
                  \vspace{-3 mm}
\end{align}
Constraint \ref{eq:dist} captures the damage containment requirement as it forces each multi-tenant transaction $t' \in T'$ to access at least two data stores, while constraint \ref{eq:capcon} ensures that the partitions can fit into an assigned data store.

%We perceive that the aforementioned optimal data partitioning with both security and performance consideration will reduce the number of affected transactions as illustrated in Figures \ref{fig:aff3}. However, the optimal data partitioning is expected to incur additional communication cost overhead as compared to the the performance-based partitioning as depicted in Figure \ref{fig:cost3}. The cost overhead is due to the synchronization of distributed transactions and the intrusion management overhead. The results of the above optimization problem can be used to guide the reengineering design of the cloud data center to achieve the best trade-off between security and performance. 

\section{Malicious Transaction Workload Benchmark}
Several benchmarks have been developed in the literature to evaluate the performance of transactional databases systems, e.g., TPC-C\footnote{http://www.tpc.org/tpcc/}. However, no benchmark has been developed to evaluate the performance of intrusion management in these systems. To address this challenge, we have developed a novel malicious transaction workload benchmark\footnote{The benchmark is available on https://bitbucket.org/mfelemban/mtb/} with the ability to generate transactional workload with several parameters and to orchestrate various attack scenarios. 

\subsection{Transactional Workload Generation}
\begin{figure}[!t]
  \centering
  \subfigure[1-to-$N$]{\label{fig:tx1n}  \includegraphics[scale=0.2]{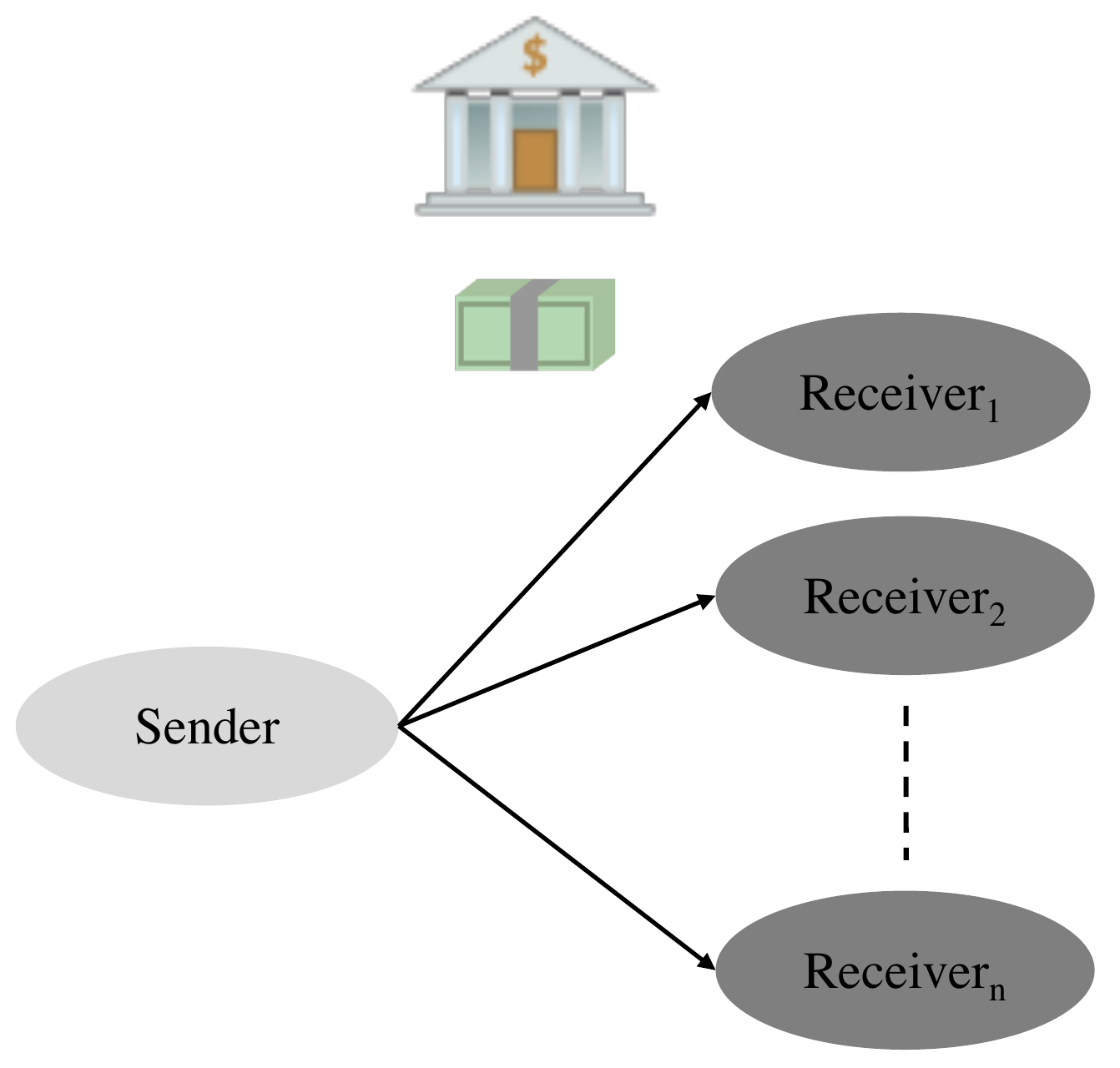}}%
  \hspace{0.2cm}
  \subfigure[$N$-to-1]{\label{fig:txn1}\includegraphics[scale=0.2]{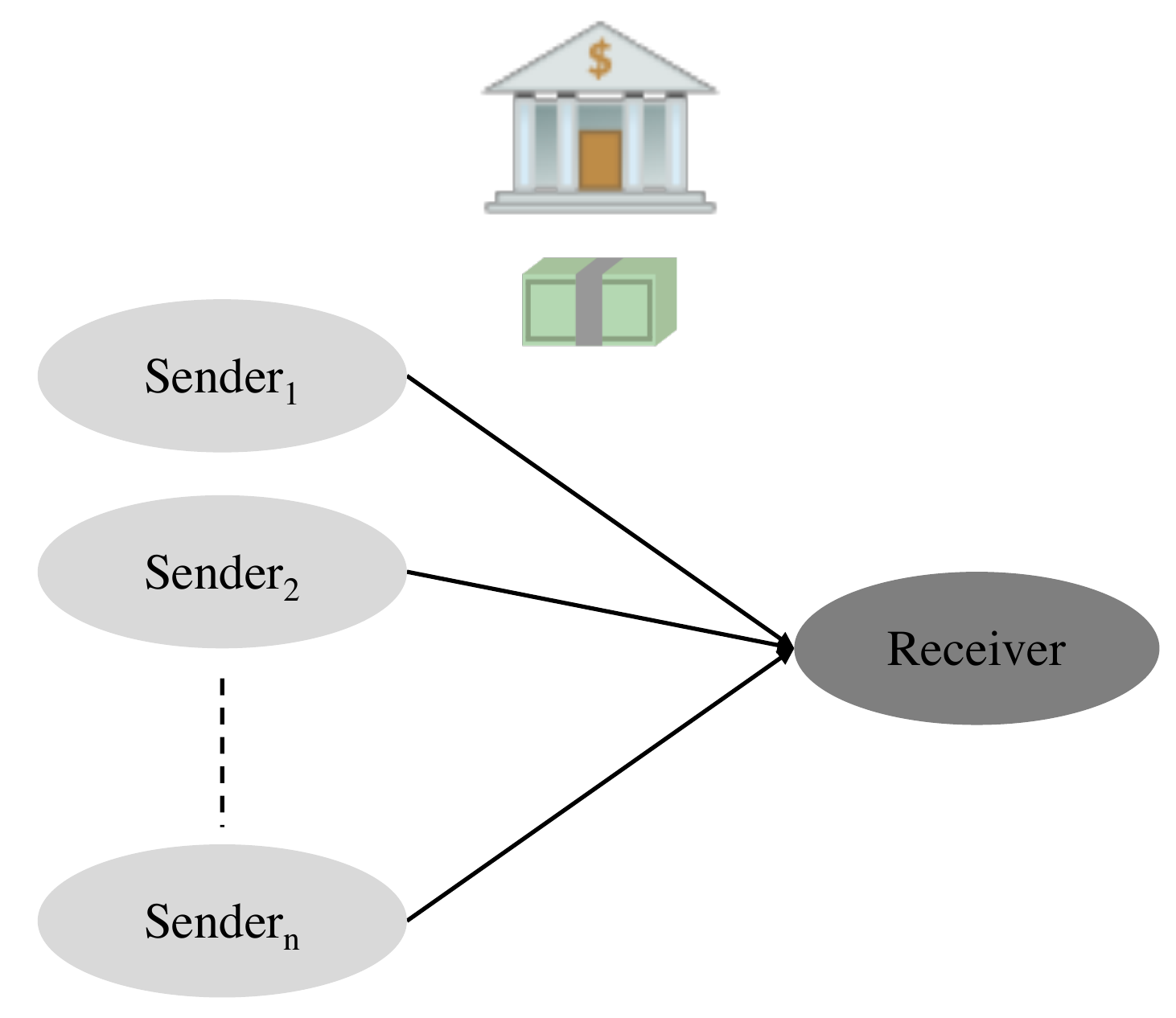}}
  \hspace{0.2cm}
  \subfigure[$N$-to-$M$ ]{\label{fig:txnm}\includegraphics[scale=0.2]{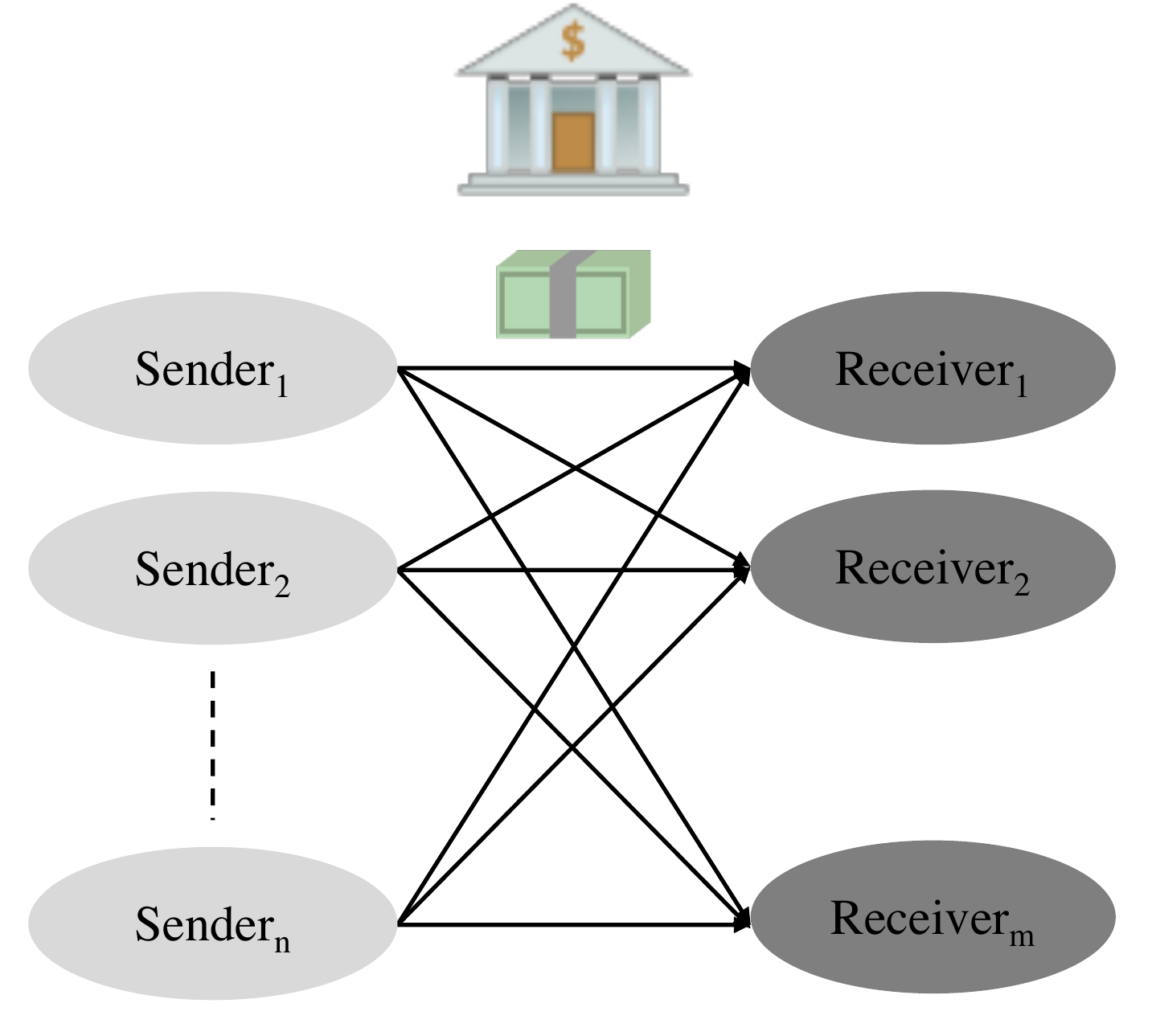}}%
  \caption{Types of money transfer transactions.}
  \label{fig:txtypes}
                    \vspace{-3 mm}
\end{figure}

The proposed benchmark simulates a banking money transfer application. In essence, the benchmark consists of a single data table, \textit{Checking}, that has two attributes: \textit{account id} and \textit{balance}. The benchmark has three types of money transfer transactions: \textit{distribute}, \textit{collect}, and \textit{many-to-many} transfer. A distribute transaction transfers money from a single account to N other accounts (a fan-out transaction); a collect transaction transfers money from M accounts to a single account (a fan-in transaction); a many-to-many transactions transfers money from many accounts to many accounts (a fan-in/fan-out transaction). A pictorial illustration of these transactions is given in Figure \ref{fig:txtypes}.

The transactional workload is generated using three parameters: the number of transactions ($n$), the degree of fan-in/fan-out ($\alpha$), and inter-transaction dependency threshold ($\beta$). The type of each transaction is chosen as distribute, collect, and many-to-many, randomly. For each transaction, the degree of fan-in/fan-out is determined using a uniform distribution with a range of [2, $\alpha$] and the amount of transferred money from the source account is chosen as a percentage of the balance with uniform distributed range of [0.01,0.1]. To simulate the transaction inter-dependency, a random graph $G$ of $n$ nodes is generated using, for example, the Erd\"os-Renyi model with $\beta$ representing the edge dependency probability. Subsequently, we map the transactions to the nodes in $G$. Two transactions are dependent if their respective nodes in $G$ are adjacent. The data objects are assigned to the transactions such that two dependent transactions are assigned shared objects. 

%In order to orchestrate the attacking scenarios, two parameters are used: the number of malicious transactions ($m$), such that $m < n$, and the malicious transactions arrival rate ($\lambda$), which dictates the intensity of the malicious transactions in the workload. The arrival rate of the transactions in the workload is assumed to be $\lambda_{Global}$. Consequently, the intensity of the malicious transaction in the malicious workload is the ratio $\frac{\lambda}{\lambda_{Global}}$. 

\subsection{Benchmark Evaluation}
\begin{figure}[t!]
    \centering
        \centering
      \subfigure[Number of affected transactions \newline with varying values of $m$ and $\lambda$.]{\label{fig:aff2}\includegraphics[width=0.5\columnwidth]{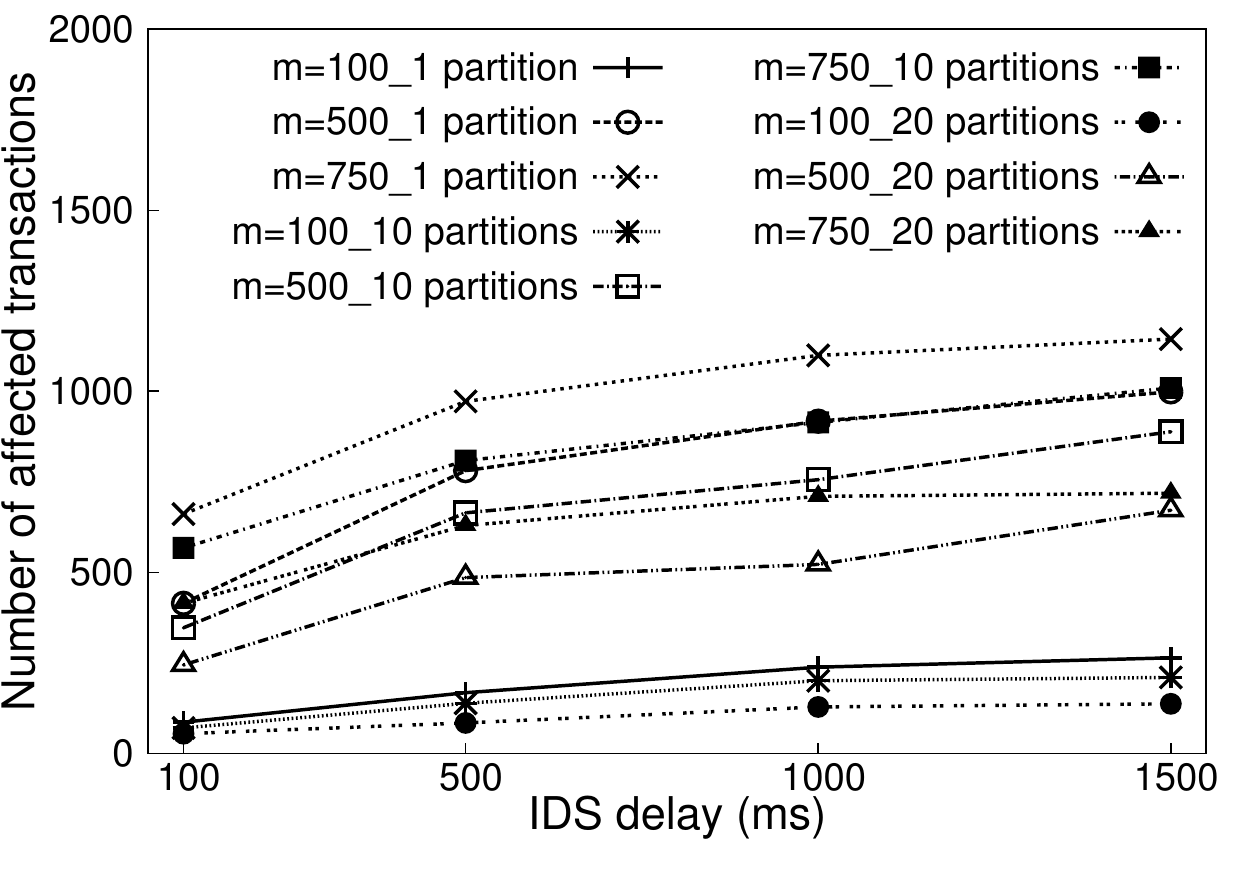}}%
  \subfigure[Average recovery time \newline with varying values of $m$ and $\lambda$.]{\label{fig:rec2}\includegraphics[width=0.5\columnwidth]{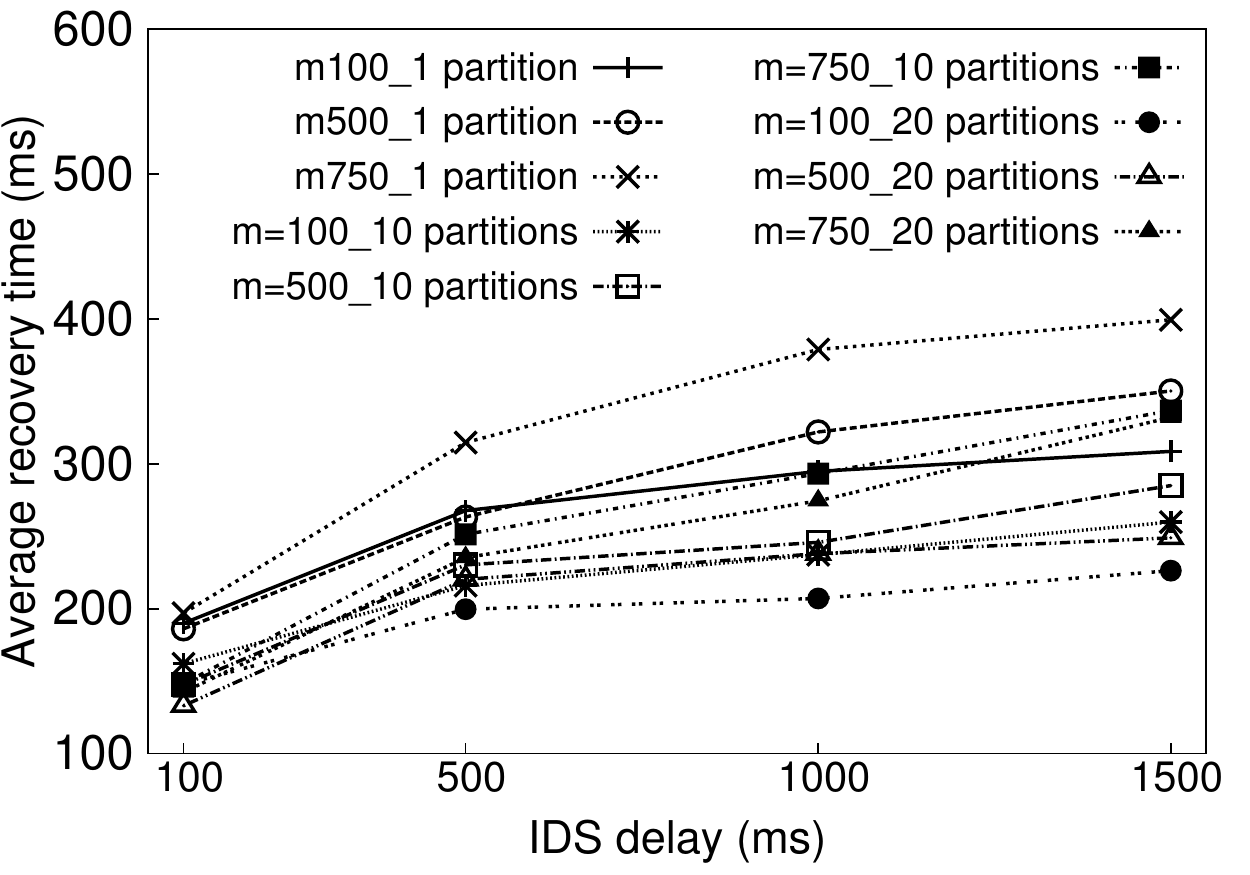}} \\
  \subfigure[Average response time \newline with varying values of $m$ and $\lambda$.]{\label{fig:res2}\includegraphics[width=0.5\columnwidth]{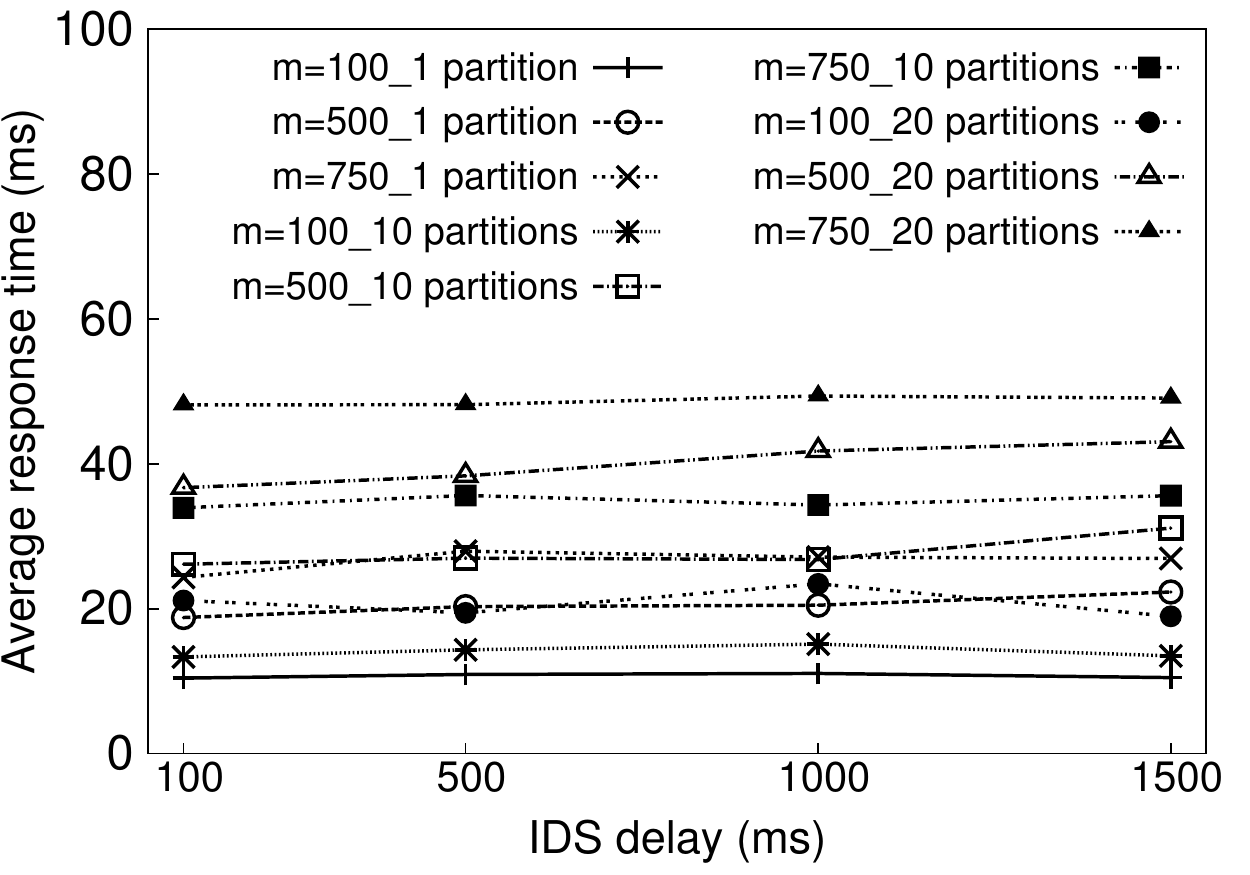}}% 
  \caption{Performance results of AIMS using various benchmarking parameters.}
  \label{fig:bench}
\end{figure}

The benchmark is implemented on an OLTP-benchmark testbed \cite{difallah2013oltp} and PostgreSQL 9.5. The balance of the \textit{accounts} in the \textit{Checking} table is initially set to \$10,000. We evaluate the performance of AIMS using three metrics: i) number of affected transactions, ii) average recovery time, and iii) average response time. The first two metrics indicate the capability of damage containment, while the third metrics indicates the availability of the database. We assume that the IDS has a detection delay of $\Delta$ $ms$. Table \ref{tab:par} shows the values of the parameters used in the experiments. The results are depicted in Figure \ref{fig:bench}.

\begin{center}
            \captionof{table}{Description of Transactions}
  \begin{tabular}{ |c|c| }
    \hline
    \small
    Parameter &  Value \\ \hline \hline
    Total number of transactions in the workload & 5000 \\ 
    $\alpha$ & 0.5 \\ 
    $\beta$ & 10 \\ 
    Transaction arrival rate & 10 Txs/sec \\
    \hline
  \end{tabular}
                \label{tab:par}
\end{center}

In this experiment, we evaluate the performance of AIMS by choosing 100, 500, and 750 malicious transactions embedded within the total workload consisting of 5000 transactions. We study the effect of IDS detection delay. It can be noticed from Figure \ref{fig:aff2} that an IDS with poor performance, i.e., with a long detection delay, can result a large number of affected transactions as compared to an IDS with a short detection delay. The reason is that as the detection delay of IDS increases, the number of dependent transactions arriving during the detection period increases. Consequently, the number of compensating transactions increases which results in a prolonged average recovery time as depicted in Figure \ref{fig:rec2}. Furthermore, a large number of affected transactions results in registering more data objects in COT. As a result, the average response time of the transactions in the workload increases because of the increase in the number of blocked transactions. The average response time is shown in Figure \ref{fig:res2}. In addition, these figures illustrate that increasing the number of malicious transactions results in increasing the number of affected transactions, average response time, and average recovery time. This is expected since increasing the intensity of attacks results in more affected transactions which can lead larger number of corrupted objects. Consequently, new arriving transactions are more likely to get blocked and hence results in an increase in the average response time as depicted in Figure \ref{fig:res2}. 

Figure \ref{fig:bench} also plots the performance of the reengineered architecture with a security-driven data partitioning and placement strategy. For this purpose, we solve the optimization problem given in Equations \ref{eq:dist}-\ref{eq:capcon} to generate 10 and 20 partitions for the data stores using a randomized heuristic. Note, the optimal solution for this type of partitioning problems is NP-Complete \cite{kumar2014sword}. In the randomized approach, we have generated a large number of random solutions and picked the one that has the smallest value of the objective function. An important observation can be made that the security-driven data partitioning and placement strategy reduces the number of affected transactions and the average recovery time as we increase the number of partitions from a non-partitioned case to a case of 10 and 20 partitions, for the same number of malicious transactions. Note, commit delays are incurred by all distributed transactions. This delay increases with the number of partitions. In particular, the commit delays associated with the distributed malicious transactions results in the reduction of the number of affected transactions as mentioned earlier. However, this reduction is achieved at the cost of increasing the average response time as depicted in Figure \ref{fig:res2}. Consequently, better results in terms of damage containment are obtained. On the other hand, as observed in Figure \ref{fig:res2} increasing the number of partitions results in larger communication overhead which subsequently leads longer average response time. In conclusion, the results show that the security-driven architecture with AIMS functionality achieves a trade-off between damage containment and performance. We expect that an improved solution for the optimization problem can further improve this trade-off. 

\section{Conclusion}
With the rapid growth of data-intensive applications, the demand on using secure and efficient cloud data centers increases. We have proposed a security-driven architecture for multi-tenant distributed transactional database in CDCs. The proposed architecture integrates the functionalities of a generic adaptive intrusion management systems with the capability to detect, respond, and recover from intrusion attacks. Furthermore, we have proposed a security-driven and performance-driven data partitioning and placement technique to support multi-tenant workload. Our future work will consider the impact of various cloud infrastructure aspects, for example networking and virtualization, on the performance of the distributed intrusion management. Further study on such aspects will lead to design CDCs with improved trade-off.

\section*{Acknowledgement}
This research was supported by the grants from Northrop Grumman Corporation and US National Science Foundation (NSF) Grant IIS-0964639.
%\section{Synchronization-based Reengineering Solution}

\vfill\eject

%\section{Author's Bio}

\noindent {\bf Muhamad Felemban} (mfelemban@purdue.edu) is an Assistant Professor in the Computer Engineering Department at KFUPM, Saudi Arabia. He has a PhD Degree from the School of Electrical and Computer Engineering at Purdue University. His research interests include security and privacy of data, IoT, and distributed systems. He is a student member of IEEE.

\noindent {\bf Anas Daghistani} (adaghist@purdue.edu) is a PhD. candidate in the School of Electrical and Computer Engineering, Purdue University. His research interests include database, distributed systems, big data management, and database security. He is a student member of IEEE.

\noindent {\bf Yahya Javed} (yjaved@purdue.edu) is a PhD student in the School of Electrical and Computer Engineering, Purdue University. His research interests include intrusion tolerance and recovery in distributed computing systems. He is a student member of IEEE.

%\noindent {\bf Anas Basalamah} (ambasalamah@uqu.edu) is an assistant professor in the Computer Engineering Department, UQU, Saudi Arabia. His areas of interest include embedded networked sensing, ubiquitous computing, participatory, and urban sensing. He is a member of IEEE.

\noindent {\bf Jason Kobes} (Jason.Kobes@ngc.com) works as a Principal Cyber Architect \& Research Scientist in Washington, DC for Northrop Grumman Corporation. Jason has over 20 years of experience concentrated in information systems design analytics, business/mission security architecture, enterprise risk management, information assurance research, and business consulting. Jason has a Master's of Science in Information Assurance (MSIA) and a Bachelor's of Science in Computer Science from Iowa State University.

\noindent {\bf Arif Ghafoor} (ghafoor@purdue.edu) is a professor in the School of Electrical and Computer Engineering at Purdue University. His research interests include: multimedia information systems, database security, and distributed computing. He is a Fellow of IEEE. 

\end{document}